\documentclass[12pt]{article}
\usepackage[body={17cm, 23cm, centered}]{geometry}
\usepackage{amsmath,amssymb,amsthm,amscd,cite,comment,amsfonts,indentfirst,color,setspace,bbold}
\usepackage{mathtext,marvosym,textcomp}
\usepackage[makeroom]{cancel}
\usepackage{array}
\usepackage{mathtools}
\usepackage{hyperref}
\usepackage{multirow}
\usepackage{longtable}
\hypersetup{colorlinks=true,linktocpage,breaklinks,
	urlcolor=blue,
	linkcolor=red,
	citecolor=blue,
    pageanchor=false
}

\usepackage[vcentermath]{youngtab}

\usepackage[mathmode,centertableaux]{ytableau}
\usepackage{tikz}
\usetikzlibrary{arrows}
\usetikzlibrary{shapes.misc}
\usepackage{tikz-cd}

\tikzset{cross/.style={cross out, draw=black, minimum size=2*(#1-\pgflinewidth), inner sep=0pt, outer sep=0pt},
	cross/.default={5pt}}
\usepackage[ normalem]{ulem}
\usepackage{tocloft}

\numberwithin{equation}{section}

\selectfont

\def\a{\alpha} 
\def\b{\beta} 
\def\g{\gamma} 
\def\d{\delta} 
\def\e{\epsilon}
\def\ve{\varepsilon} 
 
\def\h{\eta}

\def\l{\lambda} 
\def\m{\mu}
\def\n{\nu} 
 
\def\r{\rho}

\def\s{\sigma} 
\def\t{\tau}  
\def\f{\phi}

\def\L{\Lambda}

\def\p{\partial}

\def\ba{\bar{a}}

\def\fr{\frac}  \def\dt{\partial}

\def\mc{\mathcal}

\def\mD{\mathcal{D}}

\def\mF{\mathcal{F}}

\def\mL{\mathcal{L}}

\def\tf{\tilde{f}}

\def\tT{\tilde{T}}

\def\tf{\tilde{f}}

\def\XX{\mathbb{X}}
\def\RR{\mathbb{R}}

\def\Ex{\mathrm{E}}
\def\rmSL{\mathrm{SL}}
\def\rmSO{\mathrm{SO}}
\def\rmGL{\mathrm{GL}}

\def\frg{\mathfrak{g}}

\def\tfrg{\tilde{\mathfrak{g}}}
\def\frsl{\mathfrak{sl}}

\def\bas{{\mathrm{bas}}}

\def\textMathematica{\texttt{Mathematica}}

\def\ba{\begin{array}{r@{}l@{}} }
\def\ea{\end{array}}

\def\baa{\begin{array}{r@{}l  r@{} l@{} } }
\def\ea{\end{array}}

\def\EDA{\mathrm{EDA}}

\begin{document}
\renewcommand{\refname}{\begin{center}References\end{center}}
	
\begin{titlepage}
		
	\vfill
	\begin{flushright}

	\end{flushright}
		
	\vfill
	
	\begin{center}
		\baselineskip=16pt
		{\Large \bf 
		    On 10 dimensional Exceptional Drinfel'd Algebras
		}
		\vskip 1cm
			Sameer Kumar\footnote{\tt kumar.samip@phystech.edu },	
            Edvard T. Musaev\footnote{\tt musaev.et@phystech.edu}
		\vskip .3cm
		\vskip .3cm
		\begin{small}
			{\it 
			    Moscow Institute of Physics and Technology, \\ Institutskii pereulok 9, Dolgoprudny, 141700, Russia
			}
		\end{small}
	\end{center}
		
	\vfill 
	\begin{center} 
		\textbf{Abstract}
	\end{center} 
	\begin{quote}
         Based on Mubarakzyanov's classification of four-dimensional real Lie algebras, we classify ten-dimensional Exceptional Drinfeld algebras (EDA). The classification is restricted to EDA's whose maximal isotropic (geometric) subalgebras cannot be represented as a product of a 3D Lie algebra and a 1D abelian factor. We collect the obtained algebras into families depending on the dualities found between them. Despite algebras related by a generalized Yang-Baxter deformation we find two algebras related by a different Nambu-Lie U-duality transformation. We show that this duality relates two Type IIA backgrounds.
	\end{quote} 
	\vfill
	\setcounter{footnote}{0}
\end{titlepage}
	
\clearpage
\setcounter{page}{2}
	
%\tableofcontents

\section{Introduction}

String theory is a background-dependent theory meaning that dynamics of the string is defined on a fixed background of space-time fields including the metric, the dilaton, Kalb-Ramond 2-form field, and Ramond-Ramond $p$-form fields. The moduli space of these vacua appears to be highly degenerate due to duality symmetries of string theory. Some of them, such as (abelian) T-dualities are exact perturbative symmetries of the superstring partition function at all orders in $\a'$ and $g_s$ \cite{Buscher:1987sk,Buscher:1987qj,Giveon:1994fu}. This implies that physics of the string does not change if the underlying space-time background is transformed by T-duality. Given a non-abelian algebra of isometries of a string background, abelian T-duality transformation rules can be generalized to what is called non-abelian T-duality (NATD) \cite{delaOssa:1992vci}. In contrast to the abelian case NATD is not an exact quantum symmetry of the conformal theory due to problems with definition of winding modes \cite{Giveon:1993ai}. However, the NATD transformation map can be corrected to be a valid symmetry at the leading order in $\a'$ \cite{Hassler:2020tvz,Borsato:2020wwk,Codina:2020yma}. Using the notion of non-commutative currents, the non-abelian T-duality transformations can be extended to Poisson-Lie T-dualities that are symmetries of string theory in the same sense \cite{Klimcik:1995ux,Klimcik:1995dy}.  While abelian T-duality starts from a background with certain abelian isometries and preserves them, non-abelian T-duality breaks the non-abelian algebra of initial isometries naively preventing from performing the inverse transformation. The algebraic structure behind non-abelian T-duality symmetries, that is classical Drinfeld algebras, reveals that the initial isometry becomes hidden inside the algebra. More specifically classical Drinfeld algebra $\mD$ is defined in terms of Manin triple $(\mD,\frg,\tfrg)$, where $\mD$ is a Lie algebra with non-degenerate quadratic form $\h$, and $\frg$ and $\tfrg$ are subalgebras maximally isotropic with respect to the form. The algebra $\frg$ is commonly referred to as the geometric subalgebra, and is responsible for the background space, i.e. a group manifold or a coset space, while $\tfrg$ is commonly referred to as the dual algebra and it is responsible for conservation laws of the sigma model. To illustrate that, denote $f_{ab}{}^c$ and $\tf_a{}^{bc}$ as structure constants of the algebras - $\frg$ and $\tfrg$, respectively. Then, the following holds:
\begin{equation}
    \begin{aligned}[]
        [v_a,v_b]& = f_{ab}{}^c v_c,\\
        dJ_a & = \tf_a{}^{bc}J_b\wedge J_c.
    \end{aligned}
\end{equation}
Here, vectors $v_a$ define action of $G=\exp \frg$ on itself, or on a coset space as $\d x^i = v_a{}^i \e^a$, where $x^i$ denote coordinates on the group (coset) manifold. Noether currents $J_a= J_{a\, i} dx^i$ satisfy the non-commutative conservation law. When $\tf_a{}^{bc}=0$, the currents are conserved in the usual sense. Non-abelian T-duality simply maps $\frg \leftrightarrow \tfrg$, hence vanishing $\tf_a{}^{bc}$ get replaced by non-vanishing $f_a{}^{bc}$ and the conservation law becomes non-commutative. The initial isometry becomes hidden in $\frg'=\tfrg$ and is no longer manifest. In this language, the condition for classical equations of motion for the string to satisfy is simply the Leibniz identity 
\begin{equation}
[X,[Y,Z]] = [[X,Y],Z] + [Y,[X,Z]], \quad X, Y \in \mD.
\end{equation}
Here, the brackets are given by the following relations in terms of the generators $(T_a,\tT^a)=\bas \, \mD$:
\begin{equation}
    \begin{aligned}[]
        [T_a, T_b] &= f_{ab}{}^c T_c, \\
        [\tilde{T}^a, \tilde{T}^b]& = \tf_c{}^{ab} \tilde{T}^c, \\
     [\tilde{T}^a, T_b] &= \tf_c{}^{ab}\tilde{T}^c + f_{ab}{}^cT_c.
    \end{aligned}
\end{equation}
In terms of structure constants, Leibniz identity is equivalent to Jacobi identities for $f_{ab}{}^c$ and $\tf_a{}^{bc}$ along with the following mixed identity
\begin{equation}
\tf_d{}^{bc}f_{ma}{}^d + \tf_m{}^{cd}f_{da}{}^b + \tf_a{}^{bd}f_{dm}{}^c + \tf_m{}^{bd}f_{ad}{}^c + \tf_a{}^{cd}f_{dm}{}^b = 0.
\end{equation}
For a review of the algebraic construction behind Poisson-Lie T-dualities see \cite{Klimcik:1995jn}, for a review of applications of NATD see \cite{Sfetsos:2011jw,Thompson:2019ipl}, for formulation of Poisson-Lie T-dualities in the supergravity language see \cite{Hassler:2017yza,Demulder:2019bha}, and for geometric aspects see \cite{Bugden:2019vlj,Hlavaty:2020afj}.

In the most general case when both sets of structure constants are non-zero, one is able to define the so-called Poisson-Lie duality transformations. When $\dim \frg = d$, these are such maps $C_A{}^B \in O(d,d)$ that preserve the structure of classical Drinfeld double:
\begin{equation}
    T_A \to C_A{}^B T_B, \quad T_A=(T_a,\tT^a).
\end{equation}
There is a distinguished set of such transformations called Poisson-Lie (PL) T-dualities (pluralities) when the map $C_A{}^B$ relates different realization of \textbf{the same} Drinfeld algebra. The simplest example is the swapping $\frg \leftrightarrow \tfrg$. For lower dimensional Lie algebras full classification of all possible Poisson-Lie T-dualities or likewise of all equivalent Manin triples is available \cite{Hlavaty:2002kp}. This is based on classification of all possible dual algebras $\tfrg$ for each $\frg$ belonging to the Bianchi classification of three-dimensional real Lie algebras (for more on classification of Lie algebras, see for example  \cite{Popovych:2003xb}). More generally, one may have maps $C_A{}^B$ that relate different Drinfeld algebras, for example, Yang-Baxter deformations that drew interest for their preservation of the integrability of underlying sigma-model \cite{Klimcik:2008eq}.

When extending abelian T-duality symmetries by S-dualities that are non-perturbative transformations exchanging $g_s$ with $g_s^{-1}$, one arrives at U-duality transformations that are symmetries of M-theory. Speaking more concretely, U-duality is a symmetry of classical field equations of 11D supergravity compactified on a $d$-torus. These are known as Cremmer-Julia symmetries and are given by the Exceptional groups $\Ex_{d(d)}$ \cite{Cremmer:1979up,Cremmer:1997ct}. In M-theory, whose low-energy approximation is given by 11D supergravity, U-duality can be thought of as symmetries of BPS states \cite{Obers:1998fb} or in terms of a Buscher-like procedure for M2-brane wrapping a 4-torus \cite{Duff:1989tf,Duff:1990hn}. The algebraic structure behind Poisson-Lie (PL) T-dualities can be extended to the so-called Exceptional Drinfeld algebras (EDA), that include the usual abelian U-dualities (Cremmer-Julia symmetries) \cite{Sakatani:2019zrs,Malek:2019xrf,Malek:2020hpo}. Keeping the more detailed description of EDA's to the next section, we mention that these are Leibniz algebras with generators $T_A$ on which Exceptional group $\Ex_{d(d)}$ acts in the same sense as the orthogonal group O$(d,d)$ acts on generators of the classical Drinfeld double. Nambu-Lie U-dualities are then transformations that preserve the structure of the EDA. What differs these from the PL T-duality case is that there is no naturally defined analogue of the swapping $\frg \leftrightarrow \tfrg$, simply due to the following two facts: i) dimension of the geometric subalgebra $\frg$ of an EDA is never half of dimension of the EDA itself, ii) orthogonal complement of $\frg$ inside the EDA is, in general, not a Lie algebra. For this reason, searching for pairs of 11D geometries related by a Nambu-Lie U-duality is an extremely complicated task for a general EDA. At the moment few examples of such dualities between 11D backgrounds and solutions to Type II supergravity equations are known \cite{Blair:2020ndg,Blair:2022gsx}. In \cite{Musaev:2020bwm} a general procedure has been suggested similar to the natural swapping  $\frg \leftrightarrow \tfrg$ based on external automorphisms of $\Ex_{d(d)}$ group. Further it has been used to generate few examples of mutually dual backgrounds in \cite{Musaev:2020nrt}.

In this work, we elaborate further on the results of \cite{Musaev:2020bwm,Musaev:2020nrt} that in particular state that there are no non-abelian U-dualities in the defined sense between 11D background. The narrative we follow is along the lines similar to that of \cite{Hlavaty:2020pfj} where a full classification of 6D Exceptional Drinfeld doubles based on 3D geometric algebras has been presented. Starting from the classification of four-dimensional real Lie algebras \cite{mubar}, we construct all possible EDA's for a representative of each class. For each pair of such obtained EDA's we search for an $SL(5)$ transformation relating them, that would imply the existence of a Nambu-Lie U-duality between backgrounds that geometrically realize the corresponding geometric algebras $\frg$. Restricting ourselves to only such 4D real Lie algebras that do not contain a 1d (abelian) factor we find no such transformations. The restriction is motivated by the interest only in dualities between 11D background as maps from 11D$\to$IIA/B are known.

The paper is structured as follows. In the beginning of Section \ref{sec:EDA} we briefly review the construction of Exceptional Drinfeld algebras. In Section \ref{sec:ExFTGenGeom}  we discuss the geometric realization of EDA's and Nambu-Lie U-dualities. In Section \ref{sec:class}, we present classification of 10D EDA's, given the conditions stated in the preceding section and state the main results of the paper.

\section{Exceptional Drinfeld algebras}
\label{sec:EDA}

Before proceeding with the classification of 10d EDA's, let us briefly review the algebraic construction following \cite{Malek:2019xrf,Sakatani:2019zrs}. We will be focusing on the 10d case where generators of the Exceptional Drinfeld algebra $ED_4$ are collected into the 10-dimensional representation of the SL(5) group $\bas \, ED_4 = \{T_{AB}\}$, where $A,B=1,\dots,5$. Multiplication table is then given by 
\begin{equation}
T_{AB} \circ T_{CD} = \frac{1}{2}F_{AB,CD}{}^{GH}T_{GH}.
\end{equation}
The structures constants $F_{AB,CD}{}^{GH}$ are defined by the following relations 
\begin{eqnarray}
F_{AB,CD}{}^{GH} = 4F_{AB,[C}{}^{[G}\delta^{H]}_{D]} \\
F_{AB,C}{}^{D} = \frac{1}{2}\epsilon_{ABCGH}Z^{GHD} + \frac{1}{2}\delta^{D}_{[A}S_{B]C} + \frac{1}{3}\delta^{D}_{[A}\tau_{B]C} + \frac{1}{6}\delta^{D}_{C}\tau_{AB},
\end{eqnarray}
where  $\tau$ is antisymmetric and S is a symmetric tensor, while $Z^{[AB,C]} = 0$ and $Z^{AB,C} = -Z^{BA,C}$. For the algebra to be an EDA,  components of the constants $Z^{AB,C}$, $S_{AB}$ and $\t_{AB}$ under decomposition $SL(5)\hookleftarrow GL(4)$ can be defined as (up to an $SL(5)$ transformation)
\begin{equation}
    \begin{aligned}
    Z^{ab,c} &= \frac{1}{6}\epsilon^{abcd}f_{de}{}^{e} + \frac{1}{4}\epsilon^{abef}f_{ef}{}^{c}, &&   S_{5a} = f_{ab}{}^{b}, && \tau_{5a} = - \frac{1}{2}f_{ab}{}^{b}\\
    Z^{5a,b} & = \frac{1}{6}\tilde{f}_{c}{}^{abc} , &&  S_{ab} = \frac{1}{3}\Tilde{f}_{(a}{}^{cde}\epsilon_{b)cde} , && \tau_{ab} = - \frac{1}{6}\Tilde{f}_{[a}{}^{cde}\epsilon_{b]cde}  \\
    Z^{ab,5} &= -Z^{5a,b} + Z^{5b,a}  .
\end{aligned}
\label{eq:bigFcomponents}
\end{equation}
We note that $Z^{AB,C} = -Z^{BA,C}$. The constants $F_{AB,C}{}^D$ have the same structure as the embedding tensor of \cite{Samtleben:2005bp}, and in this language the above construction implies that only the geometric flux (anholonomy coefficients) and Q-flux are turned on. The former is given by the structure constants $f_{ab}{}^c$ of the geometric subalgebra $\frg$ and the latter is given by $\tf_{a}{}^{bcd}$. The algebra is Leibniz with the fundamental identity given by the quadratic relations analogous to those of 7d maximal gauged SUGRA \cite{Samtleben:2005bp}:
\begin{equation}
\label{eqn:leibniz}
2F_{AB[C}{}^{G}F_{|G|D]H}{}^{I} - F_{ABG}{}^{I}F_{CDH}{}^{G} + F_{ABH}{}^{G}F_{CDG}{}^{I} = 0.
\end{equation}
In terms of structure constants $f_{ab}{}^c$ and dual constants $f_a{}^{bcd}$, the conditions become
\begin{equation}
    \begin{aligned}
        6f_{f[a}{}^{[c}\tilde{f}_{b]}{}^{de]f} + f_{ab}{}^{f}\tilde{f}_{f}{}^{cde} - \frac{1}{3}\tilde{f}_{[a}{}^{cde}f_{b]f}{}^f = 0 \\
        \tilde{f}_{c}{}^{abc}f_{bd}{}^d = 0,\\
        f_{de}{}^{a}\tilde{f}_{c}^{bde} - \frac{1}{3}\tilde{f}_{c}{}^{abd}f_{de}{}^e = 0 \\
        \tilde{f}_{c}{}^{abg}\tilde{f}_{g}{}^{def} - 3\tilde{f}_{c}{}^{g[de}\tilde{f}_{g}{}^{f]ab} = 0 .
    \end{aligned}
\end{equation}
The last of the above equations is also referred to as the dual Jacobi condition, just as the dual conditions in the Manin triples. It describes the internal (isolated) relations between the structure constants of the dual algebra.

As in the case of Classical Drinfeld Algebra, in general, there might exist multiple equivalent choices of the geometric subalgebra $\frg$ inside an EDA. Proper generalization of the isometry condition to the case of Exceptional structures has been given in \cite{Malek:2019xrf,Sakatani:2019zrs} and can be written as follows
\begin{equation}
\epsilon^{ABCDE}T_{AB}\otimes T_{CD}\bigg|_{\frg \otimes \frg} = 0.
\end{equation}
In other words, for a given EDA, its geometric subalgebra $\frg$ is spanned by such a subset of the whole set of generators $\{T_{AB}\}$ that satisfy the above condition. For Classical Drinfel'd Double the condition is $\h^{AB} T_A\otimes T_B=0$, implying that one may, for example, take $\bas\, \frg =\{T_a\}$, or $\bas \,\frg = \{\tT^a\}$.  For EDA's, one choice is self-evident - $\bas\,\frg = \{T_{5a}\}$, while presenting an alternative choice is usually a hard task. This implies that there is no natural generalization of the Non-Abelian T-duality transformation swapping $\frg\leftrightarrow \tfrg$ in the case of EDA's, although certain progress in defining an analogue of such a  swapping has been done in \cite{Musaev:2020bwm,Musaev:2020nrt}.

\subsection{Geometric realization and dualities}
\label{sec:ExFTGenGeom}

The algebraic structure of EDA's stands behind Nambu-Lie U-dualities of supergravity solutions. These can map solutions to 11D supergravity equations into each other or into Type IIB supergravity equations. Such duality transformations map the group manifolds corresponding to different choices of the geometric subalgebra $\frg$ into each other. For more detailed and concrete algorithm of constructing mutually dual backgrounds see \cite{Musaev:2020nrt}. Below, we will briefly recall the overall  construction and highlight relations to the Exceptional Field theory (ExFT) that provides convenient variables for writing such duality maps \cite{Hohm:2013jma, Hohm:2013pua}. These are $E_{d(d)}$-covariant field theories defined in 11-dimensional space-time with an explicit split: $11=D+d$. The D-dimensional space-time is usually referred to as the external, the $d$-dimensional space is usually referred to as internal, although no compactification is assumed.  In the $d=4$ case, relevant to the present discussion, field content of the theory includes the external metric $g_{\m\n}$, ten vector fields $A_\m{}^{MN}$, five 2-form fields $B_{\m\n M}$, and 14 scalar fields parametrized by a coset element $M_{MN}\in \rmSL(5)/\rmSO(5)$. The indices $\m=0,\dots,6$ parameterize directions of the external space-time whereas the indices $M,N=1,\dots,5$ belong to the $\bf 5$ of SL(5). For more details of the construction see \cite{Musaev:2015ces}. Here we are interested in the special case where all fields transforming in irreps of SL(5) can be decomposed in terms of  matrices $E_{AB}{}^{MN}$ (generalized vielbeins) geometrically realizing an EDA. In compact notation one writes

\begin{equation}
    [E_{AB},E_{CD}] = F_{AB,CD}{}^{EF}E_{EF},
\end{equation}

where the constants $F_{AB,CD}{}^{EF}$ are precisely the structure constants of the EDA and the brackets denote the so-called generalized Lie derivative of ExFT. 

Generalized vielbeins are parametrized by fields of 11D supergravity in the $11=7+4$ split transforming as scalars under 7-dimensional diffeomorphisms. Introducing a unity matrix $M_{AB}$ compose
\begin{equation}
    M_{MN,KL} = 2E_{MN}{}^{AB}E_{KL}{}^{CD}M_{AC}M_{BD} = M_{MK}M_{NL} - M_{ML}M_{NK}.
\end{equation}
The symmetric matrix
\begin{equation}
    M_{MN} = e^{\phi } \begin{bmatrix}
        |g|^{-\fr12} g_{ij} & -V_{i} \\
        -V_{j}  & |g|^{\fr12} (1+V^{2} )\ 
    \end{bmatrix}    
\end{equation}
is then defined in terms of the 4d metric $g_{mn}$ on the group manifold, the vector $V^{m} = \frac{1}{3!}\epsilon^{mnkl}C_{nkl}$ and a scalar field  $e^{\phi} = |g_{7}|^{-\fr1{14}}$ which is the determinant $|g_{7}|$  of  external 7 dimensional space. The metric $g_{mn}$ on the group manifold is defined as usual in terms of Maurer-Cartan forms. Let $g \in G=\exp \frg$ be an element of the group $G$ whose Lie algebra is $\frg$, then 1-forms on the group manifold $g^{-1}dg \in \frg$. In components we have
\begin{equation}
    g^{-1}dg = i r_m{}^a T_a dx^m,
\end{equation}
where $x^m$ are some coordinates on the group manifold.

Given an EDA and a choice of the isotropic subalgebra $\frg$ one can explicitly construct the corresponding generalized vielbein. A step-by-step algorithm of this procedure based on constructing adjoint action of $e^h\in G$ for some $h\in \frg$ on an element of EDA can be found in \cite{Sakatani:2019jgu}. An alternative choice of the isotropic subalgebra, if exists, is related to the given one by an SL(5) transformation 
\begin{equation}
    T'_{AB} = C_A{}^CC_B{}^DT_{CD}.
\end{equation}
If this transformation respects the structure of EDA, then the alternative isotropic subalgebra is spanned by $T'_{5a}$. Structure constants of the EDA then transform as 
\begin{equation}
\label{eq:finaldualityrelation}
    F'{}_{A'B',C'}{}^{D'} = C_{A'}{}^AC_{B'}{}^BC_{C'}{}^CC_{D}{}^{D'}F_{AB,C}{}^D.
\end{equation}
Note that not any such matrix corresponds to a Nambu-Lie U-duality transformation. Indeed, one can always perform a GL(4) transformation on generators of a given algebra $\frg$ thus changing explicit realization of the corresponding EDA. Two EDAs related by such transformation then correspond to 11D backgrounds related by a coordinate transformation. Another trivial choice is
\begin{equation}
    C_A{}^B=
        \begin{bmatrix}
            \mathbf{1}_{4\times 4} & \l_{abc} \\
            0 & \mathbf{1}_{6\times 6}
        \end{bmatrix},
\end{equation}
that corresponds to simply a gauge transformation of the 3-form $C_{mnk}$. To avoid counting of EDA's related by a rotation of the basis of their isotropic subalgebras we first classify Exceptional Drinfeld algebras using classification of 4D real Lie algebras.

%%%%%%============================================================================%%%%%%%%%%%

\subsection{Classification of 10 dimensional EDA's}
\label{sec:class}

The main goal of this work is to investigate relations between 10d EDA's that correspond to Nambu-Lie U-duality transformations of 11-dimensional supergravity backgrounds. For this purpose, we start with a classification of 10D EDA's of certain class based on the classification of 4-dimensional real Lie algebras by Mubarakzyanov \cite{mubar} (for a review in English see \cite{Popovych:2003xb}). Since explicit examples of Nambu-Lie U-dualities between 11D and Type IIA/B backgrounds are known in the literature, we are interested here only in EDA's constructed on 4d real Lie algebras $\frg_4$ that cannot be decomposed into a sum $\frg_4 = \frg_4 \oplus \frg_1$, where $\frg_3$ is a 3d Lie algebra and $\frg_1$ is 1-dimensional Abelian factor. We list all relevant 4d real Lie algebras in Table \ref{tab:table1}. 

\begin{longtable}{ | r | l || r | l || r | l |}

%\centering
    \hline
    $\frg_{4,1}$ & $ \ba [T_2, T_4]&{}= T_1 \\{} [T_3, T_4] &{}= T_2 \ea$  & $\frg_{4,5}$ &$\ba [T_1,T_4]&{}=A T_1\\{} [T_2,T_4]&{} = B T_2\\{} [T_3,T_4]&{} =CT_3\\ ABC {}& \neq 0 \ea$ &  $\frg_{4,9}$& $\ba [T_2,T_3]&{}=T_1 \\{} [T_1,T_4]&{}=2A T_1 \\{} [T_2,T_4]&{}=A T_2 - T_3 \\{} [T_3,T_4]&{}=T_2 + A T_3 \\{} A&{} \geq 0 \ea$ \\
    \hline
    $\frg_{4,2}$& $\ba [T_1, T_4]&{}=\beta T_1 (\beta \neq 0) \\{} [T_2, T_4]&{}=T_2 \\{} [T_3,T_4]&{}=T_2+T_3\ea$ & $\frg_{4,6}$ &$ \ba [T_1,T_4]&{}=A T_1 \\{} [T_2,T_4]&{} = B T_2 - T_3 \\{} [T_3,T_4] &{} = T_2 + B T_3\\ A&{} >0 \ea$ & $\frg_{4,10}$&$\ba [T_1,T_3]&{}=T_1 \\{} [T_2, T_3]&{}=T_2 \\{}[T_1,T_4]&{}= -T_2 \\{} [T_2,T_4]&{}=T_1 \ea$ \\
    \hline
    $\frg_{4,3}$& $\ba [T_1,T_4]&{}=T_1 \\{} [T_3,T_4]&{}=T_2 \ea $ & $\frg_{4,7}$ & $\ba [T_2,T_3]&{} =T_1 \\{} [T_1,T_4]&{} =2T_1\\{}[T_2,T_4]&{} =T_2 \\{} [T_3,T_4]&{}=T_2 + T_3 \ea$ & $2\frg_{2,1}$& $\ba [T_1,T_2]&{} =T_1 \\{} [T_3,T_4]&{} =T_3 \ea$  \\
    \hline
    $\frg_{4,4}$& $\ba [T_1,T_4]&{}=T_1\\{} [T_2,T_4]&{}=T_1 + T_2 \\{} [T_3,T_4]&{}=T_2 + T_3\ea $ &  $\frg_{4,8}$& $\ba [T_2,T_3]&{} =T_1 \\{} [T_1,T_4]&{} =(1+\beta)T_1\\{} [T_2,T_4]&{}=T_2 \\{} [T_3,T_4]&{}=\beta T_3  \\{} \beta &{} \in [-1,1]\ea$ & & \\ 
    \hline 
\caption{Classification of 4-dimensional indecomposable real Lie algebras $\frg_{4,n}$ with $n=1,\dots,10$. The algebra $2\frg_{2,1}$ is decomposable, however does not have a $\mathfrak{u}(1)$ factor.}
\label{tab:table1}
\end{longtable}

To arrive at the corresponding classification of 10d EDA's,  we solve quadratic constraints for each class in the table above to find all possible sets of the dual structure coefficients $\tilde{f}_d{}^{abc}$. To solve the equations, we use mathematical software  \textMathematica,  that gives us all the 4 dimensional EDA's in the chosen class. We have also used GL(4) transformations inside the SL(5) duality group to set as many dual structure constants to zero as possible.  The result is listed in Table \ref{tab:table2}, where only unique combinations of indices are explicitly given in the coefficients of the underlying algebra. The rest of the indices are obtained by the antisymmetric property of the structure coefficients.  To simplify notation, we will use the symbol $c_{ij}$ to denote an arbitrary real number. The coefficients not written in the table below and not obtainable from the ones given in the table via the antisymmetric property of $\tf_a{}^{bcd}$ are to be understood as being equal to zero.

\pagebreak

\begin{longtable}{ | r | l| l | }

%\centering
    \hline 
   & \textbf{EDA}& \textbf{Structure constants $\tf^{abc}{}_d$ } \\ 
    \hline
    $\frg_{4,1}$ & 1. & $\tilde{f}_3{}^{123}=c_{11} $ \\ 
    \hline
    $\frg_{4,2}$ & 2.  & $\tilde{f}_4{}^{123} = c_{21} $\\
    \hline
    $\frg_{4,3}$ & 3. & $\tilde{f}_1{}^{123} = c_{31}, \tilde{f}_4{}^{123} = c_{32}    $ \\
    \hline
    $\frg_{4,5}$ & 4. & $\tilde{f}_4{}^{123} = c_{41} $ \\
    \hline
    $\frg_{4,6}$ & 5. & $\tilde{f}_4{}^{123} = c_{51} $ \\
    \hline
    $\frg_{4,4}$ & 6. & $\tilde{f}_4{}^{123} = c_{61} $ \\
    \hline
    $\frg_{4,7}$ & 7. & $\tilde{f}_i{}^{jkl} \equiv 0 $ \\
    \hline
    $\frg_{4,8}$ & 8. & $\tilde{f}_i{}^{jkl} \equiv 0 $ \\
    \hline
    $\frg_{4,9}$ & 9. & $\tilde{f}_i{}^{jkl} \equiv 0 $ \\
    \hline
    $\frg_{4,10}$ & 10. & $\tilde{f}_i{}^{jkl} \equiv 0 $ \\
    \hline
    $2\frg_{2,1}$ & 11. & $\tilde{f}_i{}^{jkl} \equiv 0 $ \\
    \hline 
\caption{All possible structure constants of 10d EDA's for each $\frg_{4,n}$ with $n=1,\dots,10$ and $2 \frg_{2,1}$. The numbers $c_{ij} \in \mathbb{R}$ are arbitrary constants.}
\label{tab:table2}
\end{longtable} 

Depending on particular values of the parameters of the initial four-dimensional Lie algebras and of the dual structure constants we find three essentially distinct families of Exceptional Drinfeld algebras. All algebras in the C-series have vanishing dual structure constants and hence are trivial Exceptional Drinfeld algebras. Algebras in the B-series can be represented as generalized Yang-Baxter transformations of their geometric subalgebra. This means that dual structure constants can be represented in the form

\begin{equation}
    \label{eq:dualro}
    \tf_{a}{}^{bcd} = \e^{eg[bc}\r_e f_{ga}{}^{d]},
\end{equation}
where the constants $\r_a$ satisfy the generalized Yang-Baxter equation  (gCYBE) and the unimodularity constraint
\begin{equation}
    \begin{aligned}
        f_{a[c}{}^g\r_{d]}\r_{b} - f_{b[c}{}^g\r_{d]}\r_{a}&=0,\\
        \r_{[a}f_{bc]}{}^d&=0.
    \end{aligned}
\end{equation}

Exceptional Drinfeld algebras in the A-series are neither of that, given the dual structure constant are not taken to be zero. 
The reasons for the conditions in brackets in \eqref{eq:series} are clear. The gCYBE and unimodularity constraint for $\EDA_{3}$ are proportional to $c_{31}$ and hence are satisfied when $c_{31}=0$. For $\EDA_{2,5,6}$ the constants $\r_a$ are proportional to inverse powers of $2+\b$, $\a+\b+\g$ and $\a+2\b$ respectively. Hence, if these are zero, equation \eqref{eq:dualro} can never be satisfied. The upper index $\pm$ denotes two distinct families of Exceptional Drinfeld algebras which cannot be transformed into each other by a real SL(5) transformation.

The $A_3$ and $B_3$ classes of algebras coming from EDA${}_3$ have the following subtleties. If $c_{31} \neq 0$, the algebra cannot be represents as a generalized YB transformation of it geometric subalgebra. However, in this case a $\rmGL(4)$ transformation inside the $\rmSL(5)$  group can be used to make $c_{32}=0$. This gives the $A_3$ series. On the contrary, if $c_{31}=0$, there is no such a $\rmGL(4)$ transformation, however $c_{32}$ can be transformed without changing the sign. Again, in this case, there exists a generalized YB transformation relating the EDA to its geometric subalgebra. This is the $B_3^\pm$ series.

\begin{equation}
\label{eq:series}
    \begin{aligned}
        A_1 & = \EDA_1[c_{11}\neq 0] ,   \\ 
        A_{2}^{\pm} & = \EDA_2\left[\b = -2, c_{21}=\pm 1\right], && & B_2 & = \EDA_2[\b \neq -2, c_{21}\neq 0],\\
        A_3 & = \EDA_3[c_{31}\neq 0, c_{32} = 0], && & B_3^\pm & = \EDA_3[c_{31}=0, c_{32} = \pm 1 ] \\
        A_4^\pm & = \EDA_4[\a+\b+\g = 0, c_{41} = \pm 1], && & B_4 & = \EDA_4[\a+\b+\g \neq 0, c_{41}\neq 0] \\
        A_5^\pm & = \EDA_5[\a+2\b = 0, c_{51} = \pm  1], && & B_5 & = \EDA_5[\a+2\b = 0,  c_{51}\neq 0]  ,\\
        & && & B_6 & = \EDA_6[ c_{61}\neq 0] \\
        C_{\mc{A}} & = \frg_{4,\mc{A}} \oplus_S \RR^6, \mc{A}=1,\dots,10,  \\
        C_{11}& =2\frg_{2,1} \oplus_S \RR^6 .
    \end{aligned}
\end{equation}
Here $\oplus_S$ denotes a semi-direct sum of two algebras. Note that although $T^{ab}$ generate an abelian algebra, they in general do not commute with the geometric subalgebra $\frg$ generated by $T_a$. In this case commutation relations are determined by structure constants of $\frg$.

Hence, given we are interested only in real non-trivial EDA's, we end up with 25 families of 10-dimensional EDA's 11 of which are trivial in the sense that all dual structures vanish. The B-family can also be though of as trivial in the sense that these are generated by a generalized Yang-Baxter deformation of their corresponding geometric subalgebras.

A natural question would be: whether there exists a pair of EDA's in this set that are equivalent up to an SL(5) transformation. This would mean that the same EDA can be generated by two different combinations of a 4D Lie algebra and a corresponding adjoint space to it, which together generate the EDA through the structure constants $f_{ab}{}^{c}$ and $\tf_d{}^{abc}$ respectively. In the supergravity language this would mean existence of a Nambu-Lie U-duality between 11D backgrounds geometrically realizing this pair of 4d Lie algebras. Result of our calculations is that there are no such pairs. To arrive at this statement we used \textMathematica {} software and explicitly solved equations on components of the matrix $C_A{}^B$ for each pair of 25 algebras above with no further restrictions on the coefficients. This means, that although in Table \ref{tab:table2} we list algebras as though all explicitly written dual structure constants are non-vanishing, our code does not assume that \cite{eda10}.

\subsection{Geometric realization of the duality \texorpdfstring{$A_1 \leftrightarrow A_3$ }{A1<->A3}}

There exist only one Nambu-Lie U-duality transformation understood as an SL(5) rotation of generators such that a given EDA transforms into another EDA from the above classification, that is the duality connecting $A_1$ and $A_3$ given by the following matrix
\begin{equation}
    C_A{}^B=
        \begin{bmatrix}
 0 & 0 & 1 & 0 & \frac{1}{c_{11}} \\
 -1 & 0 & 0 & 0 & 0 \\
 0 & 1 & 0 & 0 & \frac{1}{c_{11}} \\
 0 & 0 & 0 & \frac{c_{11}}{c_{31}} & 0 \\
 0 & 0 & 0 & 0 & -\frac{c_{31}}{c_{11}} 
        \end{bmatrix}.
\end{equation}
To realize the algebras in terms of 11D supergravity fields we need to construct the so-called generalized vielbein of Exceptional field theory, whose generalized anholonomy coefficients are precisely the EDA structure constants \cite{Sakatani:2019zrs,Malek:2020hpo}. To collect supergravity fields into the SL(5) generalized metric one considers the standard 7+4 KK split of the full 11D space-time metric
\begin{equation}
    ds^2_{11} = e^{-2\phi} g_{\m\n}dx^\m dx^\n + g_{mn}(dx^m + A_\m{}^m dx^\m)(dx^n + A_\n{}^n dx^\n).
\end{equation}
Here $h = \det h_{mn}$, where small Greek letters run $\m,\n=0,\dots,6$ parametrizing the so-called external space-time, while small Latin indices run $m,n=1,2,3,4$ parametrizing the so-called internal space. So far, no condition on the fields has been imposed and full dependence on all 11 coordinates is preserved. The same ansatz can be written for the 3-form field. Internal fields $g_{mn}$, $C_{mnk}$ and the scalar field $\phi$ compose the generalized metric as follows \cite{Berman:2011jh,Blair:2014zba,Bakhmatov:2020kul}
\begin{equation}
    \label{eq:metric}
    m_{MN} = e^{\f}
        \begin{bmatrix}
            g^{-\fr12} g_{mn} & -V_m \\
            -V_n & g^{\fr12}(1+V^2),
        \end{bmatrix}\in \fr{\rmSL(5)}{\rmSO(5)}\times \RR^+,
\end{equation} 
where $V^m = 1/3!\ve^{mnkl}C_{nkl}$ and capital Latin indices $M,N=1,\dots,5$ label the irrep $\textbf 5$ of $\rmSL(5)$. The corresponding generalized vielbein is defined in the usual way: $m_{MN} = E_M{}^A E_N{}^B m_{AB}$, where $m_{AB}$ is a constant matrix and $A,B=1,\dots,5$ are flat indices labeling vectors in the same irrep. The inverse $E_A{}^M$ of such defined generalized vielbeins  has the correct weight, and generalized anholonomy coefficients can be defined in the usual way
\begin{equation}
    \mL_{E_{AB}}E_C{}^M = F_{AB,C}{}^D E_D{}^M,
\end{equation}
where $E_{AB}{}^{MN} = 2 E_{[A}{}^ME_{B]}{}^N$. The generalized vielbein can be given in terms of the left-invariant 1-forms of the geometric subalgebra $\s =g^{-1}dg$ and a 1-form $\pi_m$ 
\begin{equation}
    E^M{}_A = 
        \begin{bmatrix}
            e^{\fr12} e^m{}_a & 0 \\
            -e^{\fr12}\pi_a & e^{-\fr12}
        \end{bmatrix},
\end{equation}
where $e^m{}_a$ denote the inverse of the components $e_m{}^a$ of $\s$ and the 1-form $\p_a = e_a{}^m \p_m$ satisfies
\begin{equation}
    \pi_{[a}\dt_{\vphantom{[}b}\pi_{c]} -\fr12 \pi_{[a}f_{bc]}{}^d\pi_{\vphantom{]}d}=0.
\end{equation}
This is the condition for $\pi^{abc} \equiv 1/3! \e^{abcd}\pi_d$ to define a Nambu-Lie structure. The precise form of $\pi_a$ depends on the EDA chosen and can be determined from the adjoint action of the geometric subgroup on the algebra generators $T_A$. 

The generalized Lie derivative $\mL$ of say $V^M$ along a generalized vector $\Lambda^{MN}$ is defined as follows \cite{Berman:2012vc,Musaev:2015ces}
\begin{equation}
   \d_\L V^M = \mc{L}_\L V^M = \frac{1}{2} \L^{KL} \dt_{KL} V^{M} - V^{L} \dt_{LK} \L^{M K} + \bigg(\frac{1}{4} + \lambda \bigg) V^{M} \dt_{KL} \L^{KL},
\end{equation}
where $\l$ denotes the weight. The generalized anholonomy coefficients (generalized fluxes) are the given by
\begin{equation}
    \mF_{ABC}{}^{D}  = \frac32 E_{N}{}^{D} \partial_{[A B} E^{N}{}_{C]} - E^{M}{}_{C} \partial_{M N} E^{N}{}_{[B} \delta^{D}{}_{A]} -  \frac12 E^{M}{}_{[B|} \partial_{M N} E^{N}{}_{|A]} \delta^{D}{}_{C} .
\end{equation}
The derivatives $\dt_{MN}$ and $\dt_{AB} = E_{AB}{}^{MN}\dt_{MN}$ are taken w.r.t. the coordinates $\XX^{MN}$ on the so-called extended space on which all the fields depend in the most general case. Note however, that the dependence is subject to the so-called section condition \cite{Berman:2012vc}. For us here this is not relevant as we restrict fields to depend only on the standard coordinates on the 11D space-time, which in these notations means
\begin{equation}
    \dt_{5m} = \fr{\dt}{\dt x^m}, \quad \dt_{mn} = 0.
\end{equation}

Now, after having set up the stage we turn to the results. Actual calculations are presented in the Mathematica file \texttt{Geometricrealization.nb} of \cite{eda10} and the steps to recover these results are the following. For the algebra $A_1$ we have the following left-invariant 1-forms 
\begin{equation}
    \s^{(1)} = \big(dx^1 - x^2 dx^4\big)T_1 + \big(dx^2 - x^3 dx^4\big)T_2 + dx^3 T_3 + dx^4 T_4.
\end{equation}
The Nambu-Lie 1-form has only one component $\pi^{(1)}_4 = c_{11}x^3 $ and the flux components for the corresponding generalized vielbein reproduces structure constants of $A_1$. Similarly, for $A_3$, which is the Nambu-Lie U-dual to $A_1$ we have
\begin{equation}
    \begin{aligned}
         \s^{(3)} &= \big(dx^1 - x^1 dx^4\big)T_1 + \big(dx^2 - x^3 dx^4\big)T_2 + dx^3 T_3 + dx^4 T_4,\\
         \pi^{(3)}_4& = c_{31} x^1  -e^{x^4}
    \end{aligned}
\end{equation}
for the left-invariant 1-form defining the geometric subalgebra and the only non-vanishing component of the Nambu-Lie 1-form. Note that the latter does not vanish when $c_{31}$ and $c_{32}$ are taken to zero, which has been an intentional choice to simplify further expressions. Moreover, since $A_3$ is defined as EDA$_3[c_{31}\neq 0]$ we never hit this case. The general procedure described in details e.g. in \cite{Musaev:2020nrt} in principle gives $\pi_a$ that vanishes when $c_{31}$ and $c_{32}$ go to zero, however the corresponding expression for the overall generalized vielbein and eventually the space-time metric get bulky.

To determine space-time fields one composes the generalized metric $m_{MN} = E_M{}^A E_N{}^B m_{AB}$, with a particular choice of $m_{AB}$ dictated by the equations of motion, and compares to \eqref{eq:metric}. Hence, for the algebras $A_1$ and $A_3$ we find the following geometric realizations
\begin{equation}
    \begin{aligned}
        & A_1: &&  ds_{11}^2 = ds_{(1,7)}^2+{dx}^4 \Big({dx}^4 \big(c^2_{11} \left(x^3\right){}^2-2 x^2\big)-2 {dx}^3 x^3+2 {dx}^1\Big)+2 {dx}^3 dx^2, \\
        & && C_3 = -c_{11} x^3 {dx}^2\wedge {dx}^3\wedge {dx}^4, \\
        & A_3: && ds_{11}^2 = c_{31}^{-8/5} ds_{(1,7)}+dx^4 dx^4 \left(c_{31}^2 \left(-x^1 c_{31} +e^{x^4}\right){}^2-2 x^3\right) \\
        & &&\quad -2 dx^3 \left(e^{x^4} dx^4 -c_{31} dx^1\right)+2 dx^2 dx^4, \\
        & && C_3 = -c_{31}^2 \left(-x^1 c_{31}+e^{x^4}\right) dx^1\wedge dx^3\wedge dx^4-c_{31}^{-1}{dx}^2\wedge {dx}^3\wedge{dx}^4
    \end{aligned}
\end{equation}

We see that both backgrounds have vanishing 4-form field strength $F_4=dC_3$ since $C_3$ is pure gauge in both cases. Both backgrounds are defined only by their metrics and hence are both Ricci flat due to the absence of sources. However, the corresponding Riemann tensor has non-vanishing components in both cases. Both backgrounds have the apparent Killing isometry given by $\dt_1$ in the first case and by $\dt_2$ in the second case, and can be represented in the Kaluza-Klein form
\begin{equation}
    ds_{11}^2 = e^{-\fr23 \Phi} ds_{10}^2 + e^{\fr43 \Phi}(dz - A_\a dx^\a)^2,
\end{equation}
with the coordinate $z$ chosen along the directions that realize the EDA, hence $\a=1,2,3$. The dilaton of the Type IIA theory is denoted by $\Phi$ to distinguish from the field $\phi$ determining the overall prefactor of the generalized metric. To see this for $A_1$, it is enough to perform the following coordinate transformation 
\begin{equation}
    \begin{aligned}
        x^4 & \to x^4 + c_{11} z.
    \end{aligned}
\end{equation}
For the background realizing the algebra $A_3$ we write
\begin{equation}
    \begin{aligned}
        x^1 & \to x^1 + z, \\
        x^2 & \to z(1 - c_{31}x^3),\\
        e^{x^4} & \to e^{x^4} + c_{31}z.
    \end{aligned}
\end{equation}
In both cases, these transformation render a background in the Kaluza-Klein form with non-trivial $A_\a$ and $\Phi$ that however do not depend on $z$. Hence, we conclude that the afore-found Nambu-Lie U-duality is a transformation between Type IIA backgrounds rather than a duality of spaces with full dependence on all four coordinates. This is along the lines of \cite{Musaev:2020bwm}.

\section{Discussion}

In this work we obtain a classification of 10-dimensional EDA based on the classification of 4-dimensional real Lie algebras by Mubarakzyanov \cite{mubar}. We intentionally restrict only to such 4d algebras that cannot be decomposed into a 3d algebra and a 1d abelian factor, i.e, we are interested in Nambu-Lie U-dualities between 11d backgrounds, rather than dualities between 11D and Type IIA/B solutions. More specifically, we look only at  EDAs whose isotropic (geometric) subalgebra is given by $\frg_{4,n}$ with $n = 1,...,10$ and $2\frg_{2,1}$ in terms of the Mubarakzyanov's classification. Given these restrictions, 11 possible EDAs are listed in terms of dual structure constants $\tf_a{}^{bcd}$ in Table \ref{tab:table2}. Taking into account the possibile GL(4) and generalized Yang-Baxter transformations, these split into 25 inequivalent A, B and C families of EDAs listed in \eqref{eq:series}. EDAs in the  B-series are those, that can be represented as a generalized Yang-Baxter deformation of their geometric subalgebra, EDAs in the A-series cannot be represented in such a way, while members of the C-series are trivial EDAs given by semi-direct sum $\frg \oplus_S \RR^6$ of the geometric Lie subalgebra $\frg$ and the abelian subalgebra generated by $T^{ab}$. 

The subscript $\pm$ differentiates between inequivalent families that differ by sign of a certain structure constant and there is no SL(5) transformation that related them. For example, consider the series $A_2$ and $B_2$; we know that for $\b\neq -2$ the coefficient $c_{21}$ can be put to zero by a generalized Yang-Baxter transformation that is proportional to the inverse power of $(\b+2)$. For any $\b$ a $\rmGL(4)$ transformation can change $c_{21}$, however, apparently, cannot make it vanish and hence cannot change its sign. The same can be repeated for the pairs ($A_4,B_4$) and ($A_5,B_5$). For $(A_3,B_3)$ descending from EDA$_3$ the situation is the opposite. For $c_{31}\neq 0$, there exists a $\rmGL(4)$ transformation proportional to $c_{31}^{-1}$ that turns $c_{32}$ to zero. Hence, for the $B_3^\pm$ series when $c_{31}=0$, such a transformation does not exist, however, $c_{32}$ can be turned into $+1$ or $-1$ by a $GL(4)$ transformation. Moreover $c_{32}$ can be completely removed by a generalized Yang-Baxter transformation. In this case generalized CYBE and unimodularity condition are proportional to $c_{31}$ and are hence satisfy. Certainly  algebras $B_n$ are related to $C_n$ by a generalized Yang-Baxter transformation (a particular case of Nambu-Lie U-duality). We keep them as different families of EDA's as they have geometric realizations not related to each other by a coordinate transformation.

The important question we were interested in, is whether there exists a Nambu-Lie U-duality between 11D solutions to supergravity equations beyond generalized Yang-Baxter deformations. The latter have been intensively studied beyond the algebraic approach and examples have been found e.g. in \cite{Bakhmatov:2020kul,Musaev:2023own}. Equivalently, in the algebraic language, our question can be formed as whether any of the ten Exceptional Drinfeld algebras are equivalent up to an SL(5) transformation? For that we computed the explicit form of all possible transformations between all possible pairs of EDA listed in Table \ref{tab:table2} of the form \eqref{eq:finaldualityrelation}. In our findings, we discovered that the pair of $A_1$ and $A_3$ Exceptional Drinfeld algebras can be related by a Nambu-Lie U-duality transformation that is not a YB deformation. We constructed the corresponding geometric realization of both algebras and find that this is a duality between 10D Type IIA backgrounds rather than 11D backgrounds. By this we mean that the metric along the four coordinates realizing the group manifold of the geometric subalgebra can be represented in the form of a Kaluza-Klein ansatz with no dependence on a single coordinate $z$. Hence, we conclude that there are no non-abelian U-dualities inside SL(5) Exceptional Drinfeld algebras relating 11D backgrounds. Note that this, however, does not rule out transformations between 11D and Type II backgrounds, explicit examples of which are known \cite{Blair:2020ndg,Blair:2022gsx}. Previously, in \cite{Musaev:2020bwm}, the same has been shown for transformations involving external automorphisms of the algebra $\frsl(5)$, suggested as the natural analogue of Non-Abelian T-duality transformations. Here we complete the statement.

There are further directions to extend this work. The most obvious task is to complete the classification including all 4D real Lie algebras and list sets of EDA's mutually Nambu-Lie U-dual. Less straightforwardly, we can increase the dimension of the geometric subalgebra $\frg$ by one and consider 16D Exceptional Drinfeld algebras. Unfortunately, there is no ready to use classification of 5D real Lie algebras, but certain restricted classifications are present in the literature. Some useful examples can be found in \cite{Darijani:2014ab,Andrada:2009ab,Bieszk+1997+403+424,CALVARUSO2016115}, for a review see \cite{Prieto2013AHR}.

\section*{Acknowledgments}

This work has been supported  by the Foundation for the Advancement of Theoretical Physics and Mathematics “BASIS”, grant No 21-1-2-3-1, and by Russian Ministry of Education and Science. The authors acknowledge the invaluable comments from the anonymous referee that made the presentation more clear and precise.

\providecommand{\href}[2]{#2}\begingroup\raggedright\endgroup

\end{document}